\documentclass[twocolumn]{jpsj2} 
\newcommand{\diff}{\mathrm{d}}
\newcommand{\imag}{\mathrm{Im}\,}
\newcommand{\real}{\mathrm{Re}\,}
\newcommand{\trace}{\mathrm{Tr}\,}

\title{Continuous-Time Quantum Monte Carlo Approach to
Singlet-Triplet Kondo Systems}

\author{Shintaro \textsc{Hoshino}\thanks{E-mail: hoshino@cmpt.phys.tohoku.ac.jp}, Junya \textsc{Otsuki} and Yoshio \textsc{Kuramoto}}

\inst{Department of Physics, Tohoku University, Sendai 980-8578}
\recdate{\today}

\abst{
Dynamical properties are studied numerically for a variant of the Kondo model with singlet and triplet crystalline electric field (CEF) 
levels where Kondo and CEF singlets compete for the ground state.
Using the continuous-time quantum Monte Carlo method, 
we derive the $t$-matrix of conduction electrons and dynamical susceptibilities of local electrons without encountering the negative sign problem.
When the CEF splitting is comparable to the Kondo temperature,  the dynamical response has only a quasi-elastic peak.
Nevertheless, the local single-particle spectrum shows an energy gap in strong contrast with the ordinary Kondo model.
}

\kword{continuous-time quantum Monte Carlo method, Pr skutterudite, Kondo effect, $t$-matrix, dynamical susceptibility, Pade approximation}

\begin{document}
\maketitle

\section{Introduction}
In Pr and U compounds with even number of localized $f$ electrons per site, the crystalline electric field (CEF) ground state can be a singlet.~\cite{aoki, nieuwenhuys}
Interestingly, some of these systems show the Kondo effect, while others do not.   
Hence it has long been a focus of interest how 
the competition works between the CEF singlet and the Kondo singlet under the exchange coupling with conduction electrons. 
This model is called the singlet-triplet (ST) Kondo model in the rest of the paper.
Theoretical methods for studying the ST Kondo model include
the poor man's scaling~\cite{bib3} and the numerical renormalization group (NRG)~\cite{bib4,bib5,bib_koga}. 
For Pr skutterudites, the non-crossing approximation (NCA) and the NRG has recently been used for deriving the dynamics of the  system.~\cite{bib2,bib_hattori} 
In order to clarify the details of competition not only for the ground state but also for finite temperatures,
however, more accurate theory is desirable. 

In this paper, we
derive the dynamical property of the ST Kondo
model at finite temperature using a new accurate method.
The competition between
CEF and Kondo effects is studied by
the continuous-time quantum Monte Carlo method (CT-QMC), which has been first proposed by Rubstov {\it et al}.~\cite{bib6} 
The original CT-QMC is formulated by perturbation expansion in terms of Coulomb interaction. 
Werner and Millis developed a way how the expansion is performed in terms of the hybridization between the impurity and the conduction electrons.~\cite{bib7} 
Recently, the CT-QMC method is extended to the Coqblin-Schrieffer model and the Kondo model by Otsuki {\it et al}.~\cite{bib1} 
In this paper, we apply the CT-QMC to the ST Kondo model proposed in ref. \ref{bib2_y} where the singlet-triplet states interact with the conduction-electron spins. 
It is found that the calculation 
can be performed without negative sign problem.
Hence,  the numerical results shown in this paper are highly accurate.
The single-particle excitations and the dynamical susceptibilities are obtained for the first time in the ST Kondo model.

In \S\ref{ST-Kondo}, we introduce the ST Kondo model and classify the fixed points according to the signs of two exchange interactions.
Then in \S\ref{sec_plus} we formulate the CT-QMC for the ST Kondo model,
and discuss the details of actual Monte Carlo simulations.
We introduce correlation functions which give static and dynamic susceptibilities in \S\ref{correlation}.
Numerical results for the ST Kondo model are presented in \S\ref{doublet} for the doublet ground state, and in \S\ref{singlet} for the singlet ground state.
We summarize our results in \S\ref{summary}.
In Appendix, we discuss technical aspects of extending the CT-QMC to the ferromagnetic exchange.

\section{Singlet-Triplet Kondo Model and Its Fixed Points} 
\label{ST-Kondo}


We consider the CEF singlet-triplet system interacting with conduction electrons. The singlet-triplet levels can be represented as a spin dimer with two pseudo spins $\mib{S}_1$ and $\mib{S}_2$. 
In the pseudo-spin representation of the CEF levels, the spin singlet and triplet describe the CEF singlet and triplet, respectively. 
The CEF splitting is represented by $\Delta _{\rm CEF}$.
The ST Kondo model is then written as~\cite{bib2a}
\begin{eqnarray}
{\cal H} &=& \sum_{\mib{k} \sigma } \xi _{\mib{k}} c_{\mib{k}\sigma }^\dagger c_{\mib{k} \sigma } + 2(J_1\mib{S} _1 + J_2\mib{S} _2)\cdot \mib{s} _{\rm c} \nonumber \\
 &&+ \Delta _{\rm CEF} \mib{S} _1 \cdot \mib{S} _2, \label{eqn1}
\end{eqnarray}
where $\xi _{\mib{k}}$ is the energy of the conduction electrons measured from the chemical potential, and 
the operator 
\begin{equation}
{\mib{s}}_{\rm c} = \frac{1}{2} \sum _{\sigma \sigma '} c_\sigma ^\dagger \mib{\sigma} _{\sigma \sigma '} c_{\sigma '}
\end{equation}
describes the conduction-electron spin at the origin. 
Here $c_\sigma(c_\sigma ^\dagger)$ is an annihilation (creation) operator of the conduction electron.

Using operators $\mib{X}^{\rm t} = \mib{S}_1 + \mib{S}_2$ and $\mib{X}^{\rm s} = \mib{S}_1 - \mib{S}_2$, 
the interaction term between the pseudo spins and conduction spin in eq.(\ref{eqn1}) is rewritten as
\begin{eqnarray}
{\cal H}_{\rm int } = (I_{\rm t}\mib{X}^{\rm t} + I_{\rm s}\mib{X}^{\rm s}) \cdot \mib{s}_{\rm c},
\end{eqnarray}
where $I_{\rm t} = J_1 + J_2, I_{\rm s} = J_1 - J_2$. The operator $\mib{X}^{\rm t}$ 
describes transition within triplet states, 
and $\mib{X}^{\rm s}$ connects a singlet state with triplet states.~\cite{bib2a} 
In this paper, we use the rectangular density of states for the conduction electrons:
\begin{eqnarray}
\rho (\varepsilon ) = \rho _{\rm c} \theta (D - |\varepsilon |) , \label{eqn4}
\end{eqnarray}
where $D$ is a band width and $\rho_{\rm c} = 1/2D$. We use the unit $D=1$ in numerical calculations.


In the CT-QMC method, it is difficult to treat the CEF term.
Hence we impose the condition $\Delta _{\rm CEF} = 0$. 
Even under this condition,  an effective CEF splitting arises since
two pseudo spins interact with each other through the conduction electrons.
Let us
derive the effective interaction between the pseudo spins by
applying the second-order perturbation theory to the Hamiltonian (\ref{eqn1}) with $\Delta _{\rm CEF} = 0$.
Taking the expectation value with respect to conduction electrons with the density of state (\ref{eqn4}), the effective CEF splitting in the $T\rightarrow 0$ limit is given by
\begin{eqnarray}
\tilde \Delta _{\rm CEF}= - \frac{\ln 4}{D} J_1 J_2 . \label{eqn_delta}
\end{eqnarray}
If $J_1$ and $J_2$ have different signs, $\tilde \Delta _{\rm CEF}$ is positive and the effective interaction stabilizes the CEF singlet.

\begin{figure}
\begin{center}
\includegraphics[width=6.5cm]{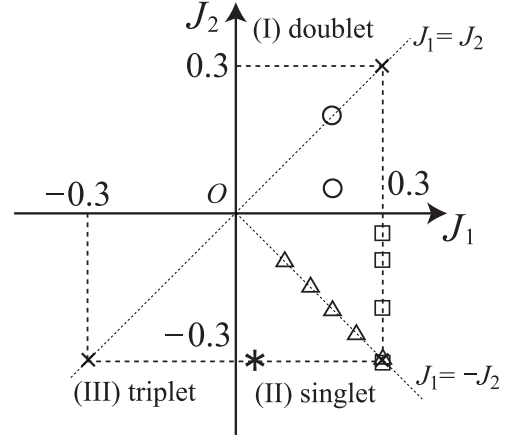}
\end{center}
\caption{Fixed points 
of the ST Kondo model for $\Delta _{\rm CEF} = 0$. 
The properties of this system are symmetric with respect to the straight line $J_1 = J_2$.
The parameters indicated by circles, squares and triangles are used later in numerical calculation. 
}
\label{fig2}
\end{figure}

We classify the parameter space into 
(I) $J_1, J_2 >0$, 
(II) $J_1>0, J_2 <0$ and 
(III) $J_1 ,J_2 <0$. 
Figure \ref{fig2} shows a schematic phase diagram for $\Delta _{\rm CEF} = 0$. The ground state of each region is (I) doublet, (II) singlet and (III) triplet. 
The residual entropy in (I) is 
understood as due to a remaining free spin after the Kondo effect compensates one of the two spins.
On the other hand, the entropies in (II) and (III) correspond to CEF singlet and triplet, respectively.
%

If a coupling constant $J_\mu$ increases toward a positive direction, the Kondo effect for the spin $\mib{S}_\mu$ becomes stronger. 
If both $|J_1|$ and $|J_2|$ increase,
the effective CEF splitting also increases according to eq.(\ref{eqn_delta}).
Hence we can see the competition between Kondo and CEF effects by 
studying representative points of
$(J_1 , J_2)$ in Fig.\ref{fig2}.
According to  NRG~\cite{bib8}, the boundary between singlet and doublet is 
in fact in the region (II) close to the $J_1$ axis.

In the region (II), 
the character of the singlet changes from CEF to Kondo as $(J_1 , J_2)$ goes away from the origin. 
From the scaling theory for the Kondo model, 
we estimate the Kondo temperature as
\begin{eqnarray}
T_{\rm K} = D \exp \left( 
- \frac{1}{2 J_1 \rho_{\rm c}} \right), \label{eqn_kondo}
\end{eqnarray}
which contains
only the antiferromagnetic coupling $J_1$. 
Since the ferromagnetic interaction is renormalized to 0 in the Kondo model, we consider eq.(\ref{eqn_kondo}) as the energy scale of the Kondo effect.
On the other hand, the characteristic energy for the CEF effect is given by eq. (\ref{eqn_delta}). 
Using these two characteristic energy values, we define the ratio:
\begin{equation}
r \equiv \frac{T_{\rm K}}{{\tilde \Delta}_{\rm CEF}}.
\label{ratio}
\end{equation}
The Kondo effect is negligible for $r \ll 1$, and 
competition against the CEF effect
arises for $r \sim 1$.

\section{Monte Carlo Procedure} \label{sec_plus}
The algorithm for the Kondo model has been developed in ref.\ref{bib1_x}. 
Here we describe how to deal with 
the ST Kondo model with $\Delta_{\rm CEF} = 0$ using CT-QMC. 
Since the original algorithm is designed for the antiferromagnetic exchange, 
we have extended the algorithm so as to be applicable also to
the ferromagnetic exchange as shown in Appendix. 

We rewrite the Hamiltonian using permutation operators as follows:
\begin{align}
{\cal H} &= {\cal H}_{\rm c} + {\cal H}_{\rm int}, \\
{\cal H}_{\rm c} &= \sum_{\mib{k}\sigma } \xi _{\mib{k}} c_{\mib{k}\sigma }^\dagger c_{\mib{k}\sigma } - \frac{J_1 + J_2}{2}\sum_{\sigma }c_\sigma ^\dagger c_\sigma, \label{eqn3} \\
{\cal H}_{\rm int} &= \sum_{\mu =1,2} J_\mu \sum_{\sigma \sigma '} X_{\sigma \sigma '}^\mu (c_{\sigma '}^\dagger c_\sigma - \alpha _\mu \delta _{\sigma \sigma '}) 
+ \sum _{\mu} \alpha _\mu J_\mu, \label{eqn_interaction}
\end{align}
which form is suitable to apply the CT-QMC.
Here $X_{\sigma \sigma '}^\mu $ is the operator that changes the $\mu $-th pseudo-spin state from $\sigma '$ to $\sigma$. 
The parameter $\alpha _\mu$ is introduced to avoid the negative sign configuration (see Appendix for detail). 
We take $\alpha _\mu$ to be 0 for the ferromagnetic coupling and 1 for the antiferromagnetic coupling.
The constant term $\sum _{\mu} \alpha _\mu J_\mu$ may be neglected. The partition function $Z=\trace e^{-\beta {\cal H}}$ is 
factorized as 
$Z = Z_1 Z_{\rm c}$ with $Z_{\rm c} = {\rm Tr}_{\rm c}\  e^{- \beta {\cal H}_{\rm c}}$. 
We obtain $Z_1$  as
\begin{align}
Z_1 &= \int {\cal D}q \  W(q), \\
\int {\cal D}q  &= \sum_{k=0}^{\infty } \int^{\rm ordered} \hspace{-8mm} \diff \tau _1 \cdots \diff \tau _k \sum_{ \{ \mu _k \} }\sum_{\{ \sigma _k \}} \sum_{ \{ \sigma ' _k \} } , \label{eqn2}
\\
W(q) &= J_{\mu _1} \cdots J_{\mu _k} \  s \prod _{\sigma } \det D_\sigma ^{(k_\sigma)} \nonumber \\
&\times\prod _{\mu } \trace [T_\tau X^{\mu {\rm I}}_{\sigma _1 \sigma _1 '}(\tau _1) \cdots X^{\mu {\rm I}}_{\sigma _{k_\mu } \sigma _{k_\mu } '}(\tau _{k_\mu }) ] , \label{eqn_weigh} \\
q &= (k, \{ \tau _k \}, \{ \mu _k\} , \{ \sigma _k \}, \{ \sigma _k ' \} ), 
\end{align}
where `ordered' in eq.(\ref{eqn2})
means that the configuration $\{ \tau _k \}$ is aligned as $\beta > \tau _k > \cdots > \tau _2 > \tau _1 \geq 0$.
The notation $\rm{I}$ means the  interaction picture 
for  the time evolution. 
The order $k$ of the expansion 
consists of  components 
$k_\sigma$ and $ k_\mu $ with the relation
$\sum _\sigma  k_\sigma = \sum _{\mu } k_\mu  =  k$. 
The suffices $\sigma$ and $\mu$ correspond to
a set of operators $c_\sigma ^\dagger c_\sigma$ and to $ X ^\mu$, respectively.

The conduction-electron operators are grouped by their spin indices.
Using Wick's theorem, the spin $\sigma $ conduction-electron part is represented as determinant of $k_\sigma \times k_\sigma $ matrix $D_\sigma ^{(k_\sigma)}$ whose elements are composed by the free conduction-electron Green function $g(\tau )$. 
A random walk in the configuration space $\{ q \}$ with the weight $W(q)$ enables us to 
perform the integral (\ref{eqn2}). 
When $k_\sigma$ increases, it takes much time to update the matrices of the conduction-electron parts in numerical calculations.
 Note that the most important value of $k_\sigma$ depends on the strength of the interaction $J_\mu $. 
Since two pseudo spins interact with conduction electrons,
 the calculation becomes heavier than the Kondo model, which has only one local spin. We also note that the antiferromagnetic coupling makes $k_\sigma$ larger than the ferromagnetic coupling.

\begin{figure}
\begin{center}
\includegraphics[width=7.5cm]{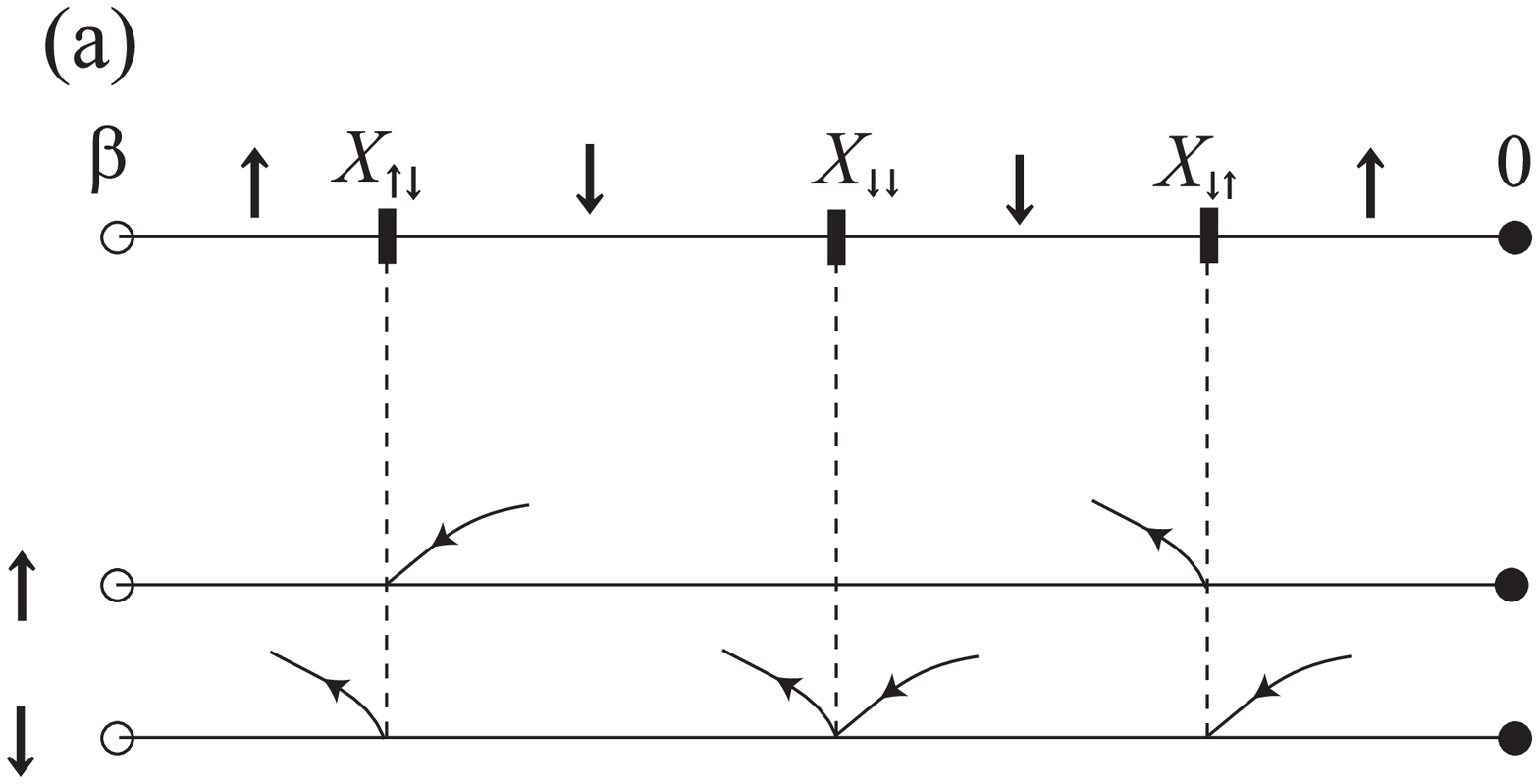}
\includegraphics[width=7.5cm]{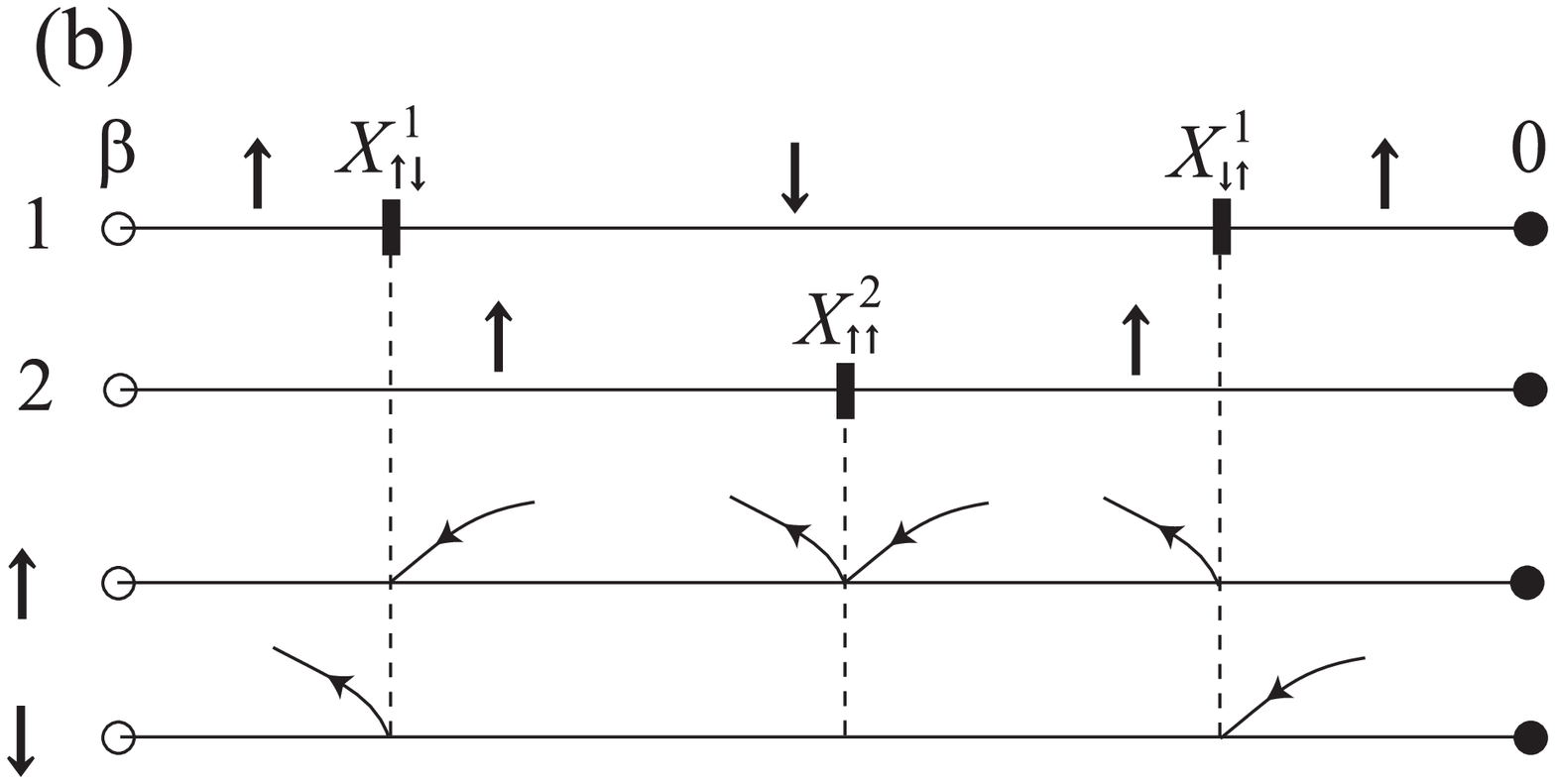}
\end{center}
\caption{Diagram of a configuration $q$ with $k_1=2$, $k_2=1$, $k_\uparrow =2$, $k_\downarrow =1$
for (a) the Kondo model and (b) the ST Kondo model.
The $X$-operators act on
the local state, while 
the incoming and outgoing arrows denote the annihilation and creation of conduction electrons, respectively.}
\label{fig1}
\end{figure}

We represent the configuration $q$ by a diagram as in ref. \ref{bib1_x}. 
Figure \ref{fig1} compares
the diagram for the Kondo model and the ST Kondo model. The ST Kondo model has two pseudo-spin components labeled by 1 and 2. 
In order to change a configuration in Fig. \ref{fig1}(a), we add a new operator $X_{\sigma \sigma '}$ to the local spin configuration, and $c_{\sigma '} ^\dagger c_\sigma$ to the conduction parts. 
In Fig. \ref{fig1}(b), on the other hand,  we first choose either $\mu = 1$ or $2$ and then add operators as in Fig. \ref{fig1}(a). Thus, the updating process in the ST Kondo model is the same as in the Kondo model except for additional choice of a pseudo-spin component.

In the simulation of the ST Kondo model, we have observed negative weight configurations at the rate of about $10^{-4}$ at low temperature, which is to be compared with $10^{-7}$ in the Kondo model. 
The occurrence here is largely due to rounding errors, since it depends on the numerical treatment of the Green-function determinant $\det D_\sigma ^{(k_\sigma)}$.
The weight is small enough to perform the simulation accurately.

\begin{figure}
\begin{center}
\includegraphics[width=7.5cm]{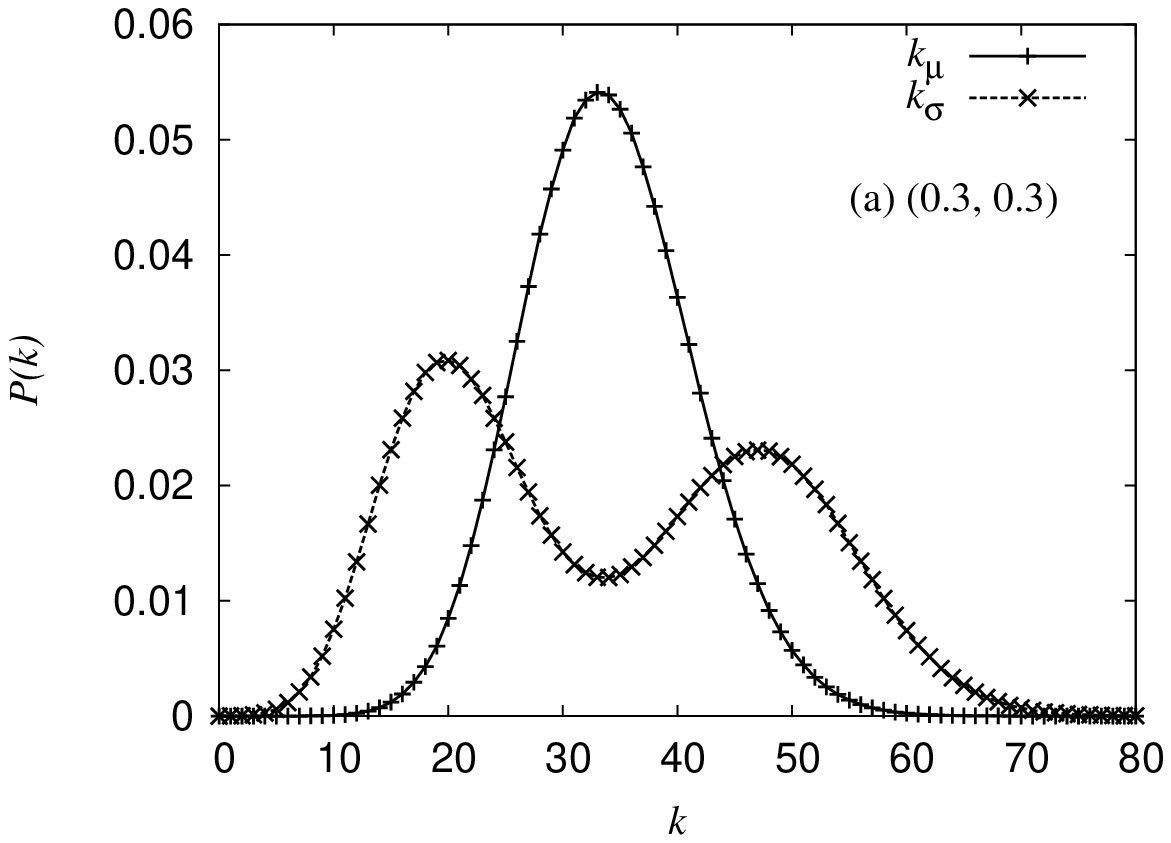}
\includegraphics[width=7.5cm]{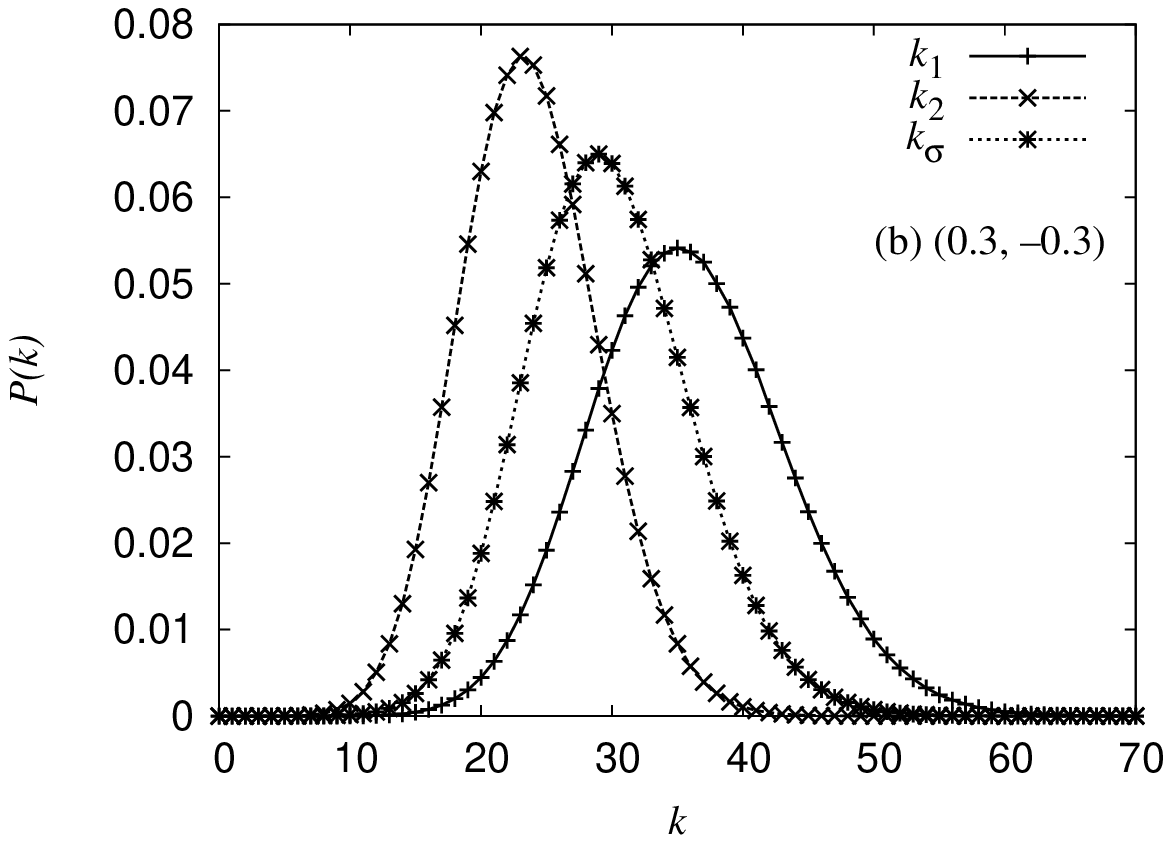}
\includegraphics[width=7.5cm]{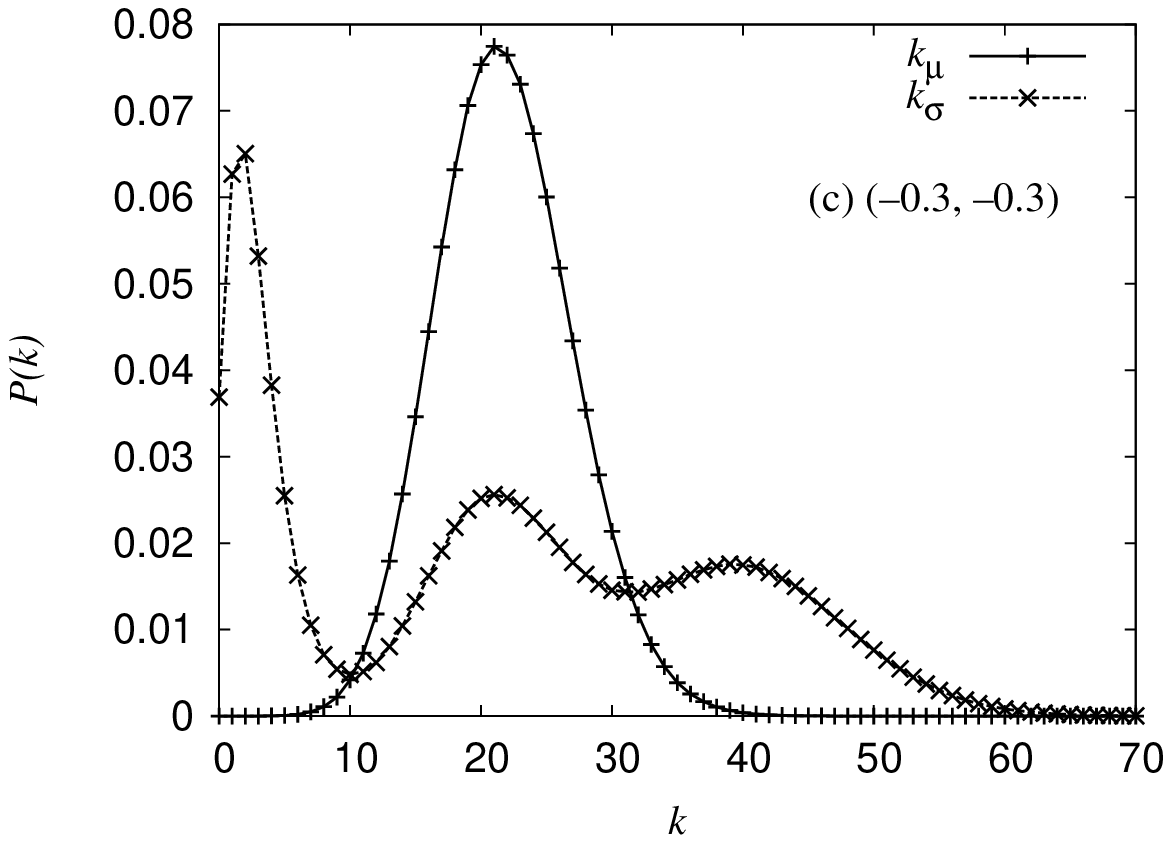}
\end{center}
\caption{The probability distribution $P(k)$
at $T=0.01$ in (a) region (I), (b) region (II), and (c) region (III). 
The exchange parameters are shown as crosses in Fig. \ref{fig2}.}
\label{figi}
\end{figure}
Figure \ref{figi} shows the probability distribution with respect to $k_\mu $ and $k_\sigma $
defined by eq.(\ref{eqn_weigh}). 
In the region (I), as shown in (a) with $(J_1, J_2)=(0.3,0.3)$, the distribution of $k_\sigma $ has two peaks, which are caused by spin fluctuation; 
if an $\uparrow $-spin is on the left peak in a snapshot of the Monte Carlo simulation, another spin with $\downarrow$ is on the right peak. 
At low temperature, these two peaks tend to separate completely. In this case, we cannot move around all the configuration space $\{ q \}$, since each spin is trapped on the different peak during the simulation. 
We can resolve this problem by introducing a sweeping procedure that flips all the pseudo-spin states. This update can be done safely
because it does not change the value of $W(q)$ in the ST Kondo model. 
On the other hand in the region (II), as shown in (b) with $(J_1, J_2)=(0.3,-0.3)$, the probability distribution has only a single peak. Hence we do not need the sweeping procedure. 

In the region (III), as shown in (c) with $(J_1, J_2)=(-0.3,-0.3)$, the probability distribution with respect to $k_\sigma$ has triple peaks due to fluctuation among three components. The number of degrees of freedom means that the triplet states are 
stabilized
against the singlet state by the effective CEF splitting. 
In the Monte Carlo simulation, presence of an $\uparrow $-spin on the left peak means presence of a $\downarrow $-spin on the right peak, which is the same as in the region (I). 
On the other hand, if an $\uparrow $-spin is on the central peak, then another spin with $\downarrow $ is also on the central peak. 
At low temperature, these three peaks separate completely. We cannot move around all the configuration space even by the sweeping spin flip because of the central peak. 
For a simulation in the region (III), therefore, we need another procedure to mix the central peak and the side peaks.  
Physically, the region (III) is less interesting because both exchange couplings $J_1, J_2$ renormalize to zero.  Hence we do not consider this region in the rest of the paper.

In the Monte Carlo simulation, we estimate statistical errors from 20 bins of data. To obtain dynamical quantities on the real-frequency axis, we have performed analytic continuation using the Pad$\acute {\rm e}$ approximation. 
Although the Pad$\acute {\rm e}$ approximation does not take statistical errors into account, 
the data of the CT-QMC are accurate enough to obtain reliable dynamics.
This aspect has already been demonstrated  in ref.\ref{bib1_x}. 
In particular, at low temperature the Pad$\acute {\rm e}$ approximation well reproduces the spectrum, since the data interval $2\pi T$ 
on the imaginary axis becomes narrow.

The ST Kondo model has the particle-hole symmetry 
in the single-particle spectrum. 
It then follows that the $t$-matrix in the Matsubara frequency domain should be pure imaginary. 
We have imposed the condition before
analytic continuation, by neglecting tiny real part arising from statistical errors.
As a result,  the spectrum maintains the particle-hole symmetry.

\section{Correlation Functions} 
\label{correlation}
As the most fundamental quantity in the system,  the imaginary time correlation function is discussed now.  
The correlation function has a label of pseudo spins, and is defined by
\begin{eqnarray}
\chi ^{\mu \nu}_{\sigma \sigma '}(\tau ) = \langle \tilde n_{\mu \sigma} (\tau ) \tilde n_{\nu \sigma '} \rangle ,
\end{eqnarray}
where $n_{\mu \sigma} = X_{\sigma \sigma }^\mu $ is the number operator of the $\mu $-th pseudo spin with a spin $\sigma$,
and tilde means deviation from the mean number: $\tilde n_{\mu \sigma} = n_{\mu \sigma} - \langle n_{\mu \sigma} \rangle $. 
Similarly we define a pseudo-spin correlation function by $\chi ^{\mu \nu }_{\rm M} (\tau) = \langle {\tilde S^z}_\mu (\tau ) {\tilde S^z}_\nu \rangle $ where $S_\mu ^z = \sum_{\sigma } \sigma n_{\mu \sigma}$ is the $\mu $-th pseudo-spin magnetic moment ($\sigma = \pm 1/2$). 
We introduce a correlation function
$\chi _{\rm M} ^{ \mu \nu}$ 
in terms of $\chi _{\sigma \sigma '} ^{\mu \nu}$ as
\begin{eqnarray}
\frac{\chi _{\rm M}^{\mu \nu } (\tau ) }{C} = \sum_{\sigma }[\chi ^{\mu \nu }_{\sigma \sigma }(\tau ) - \chi ^{\mu \nu }_{\sigma \bar \sigma }(\tau )],
\end{eqnarray}
where $C=1/4$ is the Curie constant, 
and $\bar\sigma \equiv -\sigma$. 
The correlation function has the symmetry 
$\chi_{\rm M} ^{\mu \nu} = \chi_{\rm M} ^{\nu \mu}$. 
The static susceptibility is obtained by integrating the correlation function from $0$ to $\beta$.
With use of $\mib{X}^{\rm t}$ and $\mib{X}^{\rm s}$, we define related correlation functions as follows:
\begin{eqnarray}
\chi_{\rm t,s} (\tau ) = \langle X_z^{\rm t,s} (\tau ) X_z^{\rm t,s} \rangle = \chi _{\rm M} ^{11} (\tau ) + \chi _{\rm M} ^{22} (\tau ) \pm  2 \chi _{\rm M} ^{12}(\tau ).
 \label{cor_ts}
\end{eqnarray}

The physical magnetic moment $J_z$ is given by $J_z = \sum_{\mu } a_\mu  S ^z _\mu $ 
where the coefficient $a_\mu$ depends on wave function of the local states.~\cite{bib8} 
Then the magnetic correlation function $\chi _J$ is represented by $\chi _J(\tau ) = \sum_{\mu \nu } a_\mu  a_\nu  \chi _{\rm M}^{\mu \nu } (\tau )$. 
In this paper, we show results for susceptibilities $\chi _{\rm M}^{\mu \nu }$ separately
in order to see responses of each pseudo spin.
Other multipoles like a quadrupole can be also written in terms of $\mib{S}_1$ and $ \mib{S}_2$.

The projection operator $P_{\rm s}$ onto the singlet state is given by $P_{\rm s} = - \mib{S}_1 \cdot \mib{S}_2 + 1/4$, and the triplet projection is given by $P_{\rm t} = 1 - P_{\rm s}$. 
We 
can derive the singlet occupation rate with use of the correlation function $\chi_{\rm M}^{12}$. 
In the isotropic system, the singlet occupation rate is given by
\begin{equation}
\langle P_{\rm s} \rangle = \frac{1}{4} 
- 3\chi _{\rm M} ^{12}(\tau =0).
\label{singlet_occ_rate}
\end{equation}
In the high-temperature limit, we obtain $\langle P_{\rm s} \rangle = 1/4$ from eq.(\ref{singlet_occ_rate}), since there is no correlation between different pseudo spins in this limit.

\begin{figure}
\begin{center}
\includegraphics[width=7.5cm]{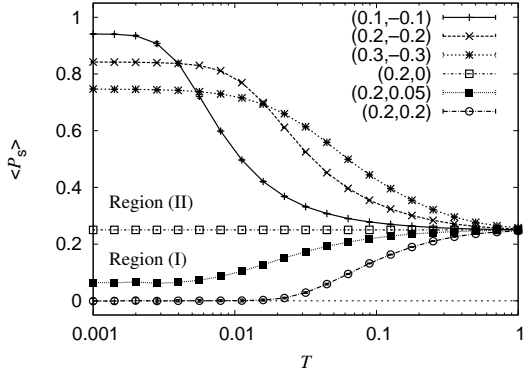}
\end{center}
\caption{Singlet occupation rate $\langle P_{\rm s} \rangle $ as a function of
temperature. 
The parameters $(J_1 , J_2)$ are shown as circles and triangles in Fig. \ref{fig2}.}
\label{fig_sing}
\end{figure}

It is convenient to characterize the 
CEF states in terms of the singlet occupation rate $\langle P_{\rm s} \rangle$. 
Figure \ref{fig_sing} shows $\langle P_{\rm s} \rangle$
for the region (I) and (II), with contrasting dependence on  temperature.
The value of $\langle P_{\rm s} \rangle$ 
changes most significantly at temperatures corresponding to the 
the effective CEF splitting. 
In the case of $(J_1,J_2)=(0.2,0.2)$,  the singlet occupation rate 
tends to 0 due to the relation 
$I_{\rm s} = J_1 - J_2=0$. 
Namely, only the triplet gains the interaction energy which is always negative in second-order.
On the other hand, with 
$(J_1,J_2)=(0.2,0.05)$, the singlet state also participate in the ground state to give $\langle P _{\rm s} \rangle \neq 0$.

In the region (II), 
$\langle P_{\rm s} \rangle $ comes close to unity 
with $r\ll 1$ as defined by eq.(\ref{ratio}).
For example, 
in the case of $(J_1,J_2)=(0.1,-0.1)$ 
we obtain $r = 0.0033 \ll 1$,  
and the CEF effect is dominant over the Kondo effect.
For larger $J_1=-J_2$ in the region (II), 
the Kondo effect becomes important, and the singlet occupation rate 
$\langle P_{\rm s} \rangle $ decreases.
For example,  the case $(J_1,J_2)=(0.3,-0.3)$ gives $r = 0.29 \sim 1$, 
and $\langle P_{\rm s} \rangle $ does not tend to unity even at low temperature. 

In the following, we discuss the regions (I) and (II) in more detail.

\section{Doublet Ground State by Underscreened Kondo Effect}
\label{doublet}

Let us first concentrate on  the region (I) where the underscreened Kondo effect occurs.
As a typical example, we take the case $J_1=J_2=J>0$ where
the interaction Hamiltonian is written as
\begin{equation}
{\cal H}_{\rm int} = J(\mib{S}_1 + \mib{S}_2) \cdot \mib{s}_{\rm c} .
\end{equation}
In the strong-coupling limit, we can neglect the kinetic energy of conduction electrons.
Then the ground state $|\rm g\sigma\rangle$ is a doublet composed by linear combination of 
$|S_1^z + S_2^z,s_{\rm c}^z \rangle$ where only
the triplet ($S=1$)
part enters in $S_1^z + S_2^z$. 
Namely we obtain
\begin{equation}
|\rm g\uparrow\rangle = \sqrt{\frac 23}|1,-\frac 12\rangle
-\sqrt{\frac 13}|0,\frac 12\rangle,
\end{equation}
and the time-reversal partner $|\rm g\downarrow\rangle$
with the corresponding energy $E_{\rm g} = -J$.
We have the relation
$\chi_{\rm M} ^{11} =
\chi_{\rm M} ^{22}$ because of the condition $J_1 = J_2$. 
The squared effective moment is given by 
\begin{equation}
T \chi_{\rm t} = 
|\langle{\rm g}\uparrow |S_1^z + S_2^z| {\rm g} \uparrow\rangle |^2 =
4/9 \simeq 0.44
\label{strong}
\end{equation}
for the ground state. 

\begin{figure}
\begin{center}
\includegraphics[width=7.5cm]{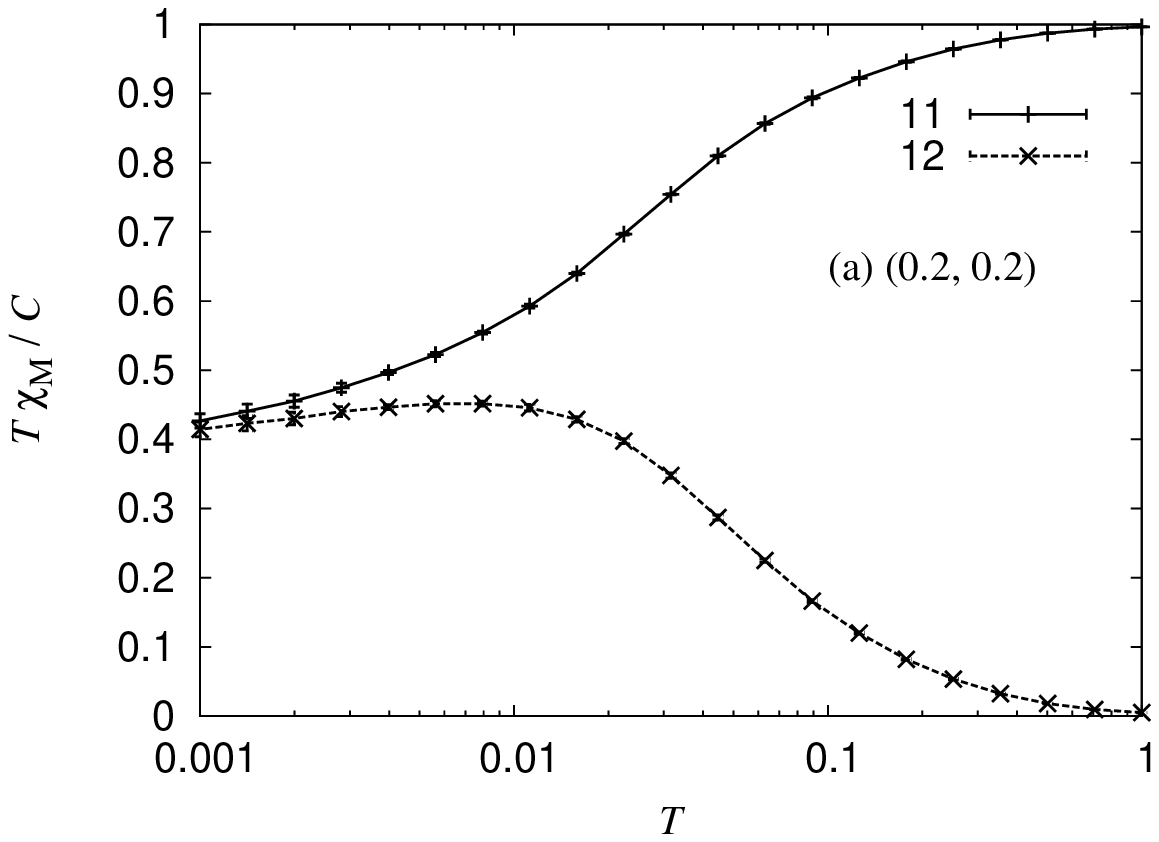}
\includegraphics[width=7.5cm]{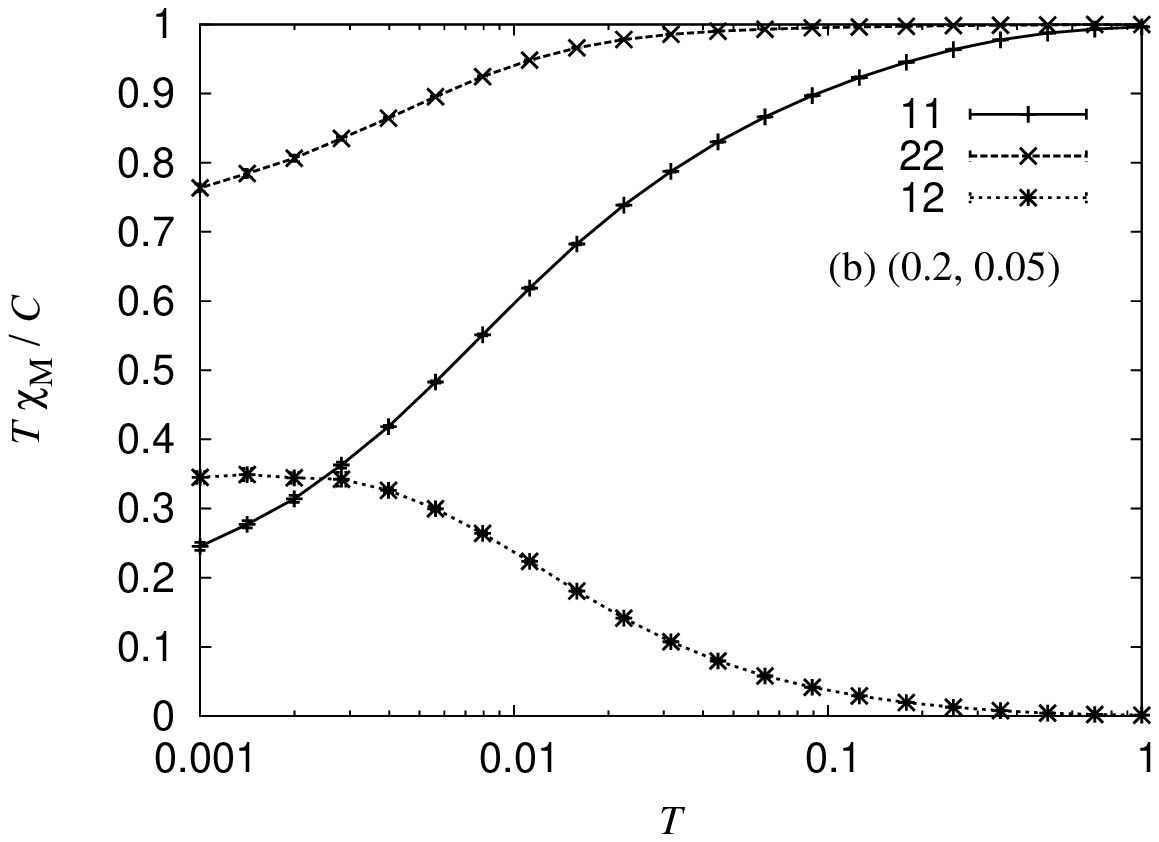}
\end{center}
\caption{
$T\chi _{\rm M} ^{\mu \nu}/C$
as a function of temperature in the region (I). 
The symbols 11, 22 and 12 mean $\chi_{\rm M}^{11}$, $\chi_{\rm M}^{22}$ and $\chi_{\rm M}^{12}$ respectively. Parameters are shown as circles in Fig. \ref{fig2}. }
\label{static}
\end{figure}

Figure \ref{static} shows the numerical results for
$T\chi _{\rm M} ^{\mu \nu}/C$ in the region (I).  The property
$\chi _{\rm M} ^{12} > 0$ indicates the ferromagnetic correlation between the pseudo spins $\mib{S}_1$ and $\mib{S}_2$. 
At high temperature, the result tends to 
the Curie law $T\chi _{\rm M} ^{11}/C = 1$. 
The effective moments seem to become constant at sufficiently low temperature. Note that the value is unity if we have a free spin with $S=1/2$.
%
In Fig. \ref{static}(a) with $J_1=J_2=0.2$,  
both $T\chi _{\rm M} ^{11}$ and $T\chi _{\rm M} ^{12}$ tend to the same value in the low temperature limit. 
This is because the condition $\chi _{\rm M}^{11} = \chi _{\rm M}^{12} = \chi_{\rm t}/4$ is satisfied at sufficiently low temperature due to the absense of $I_{\rm s} = J_1 - J_2$.
If there is no Kondo effect, we expect $T \chi _{\rm t} = 2/3 \simeq 0.67$  
associated with the triplet ground state. 
On the other hand, the strong coupling limit gives $T \chi _{\rm t}\sim 0.44$ as given by eq.(\ref{strong}).
The computed value
$T \chi _{\rm t} \simeq 0.4$ in the low temperature limit is 
close to the strong coupling limit, 
and is far from the values for a free spin with either $S=1/2$ or $S=1$.
On the other hand, if we have $J_1\neq J_2$ as shown in Fig. \ref{static}(b),  we no longer have the condition $\chi _{\rm M}^{11} = \chi _{\rm M}^{12}$.  In this case, the effective moment tends to another value which depends on the ratio $J_1/J_2$.

\begin{figure}
\begin{center}
\includegraphics[width=7.5cm]{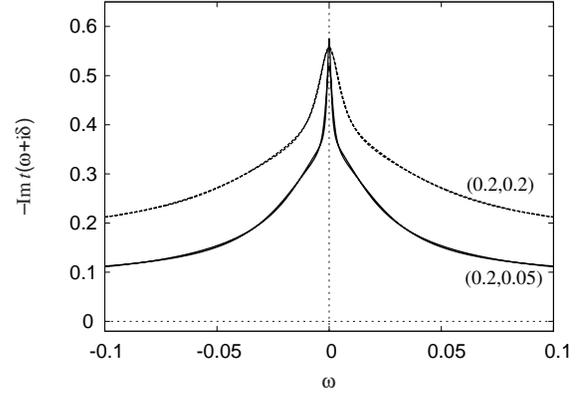}
\end{center}
\caption{$t$-matrix at $T=0.001$ in the region (I). 
Two lines corresponding to up and down spins
are given for each parameter set $(J_1 , J_2)$.
The good agreement of the two lines indicates the accuracy
of Pad$\acute {\rm e}$ approximation. 
The parameters $(J_1 , J_2)$ are 
shown as circle points in Fig. \ref{fig2}.}
\label{t_mat}
\end{figure}

The single-particle spectrum of the underscreened Kondo system is also interesting but
has not been investigated so far to our knowledge.  
The relevant quantity is
$-\imag t_\sigma (\omega +{\rm i} \delta )$ 
where $t_\sigma (z)$ is the impurity $t$-matrix of spin $\sigma$. 
We derive $-\imag t_\sigma (\omega +{\rm i} \delta )$ following the procedure of ref.\citen{bib1}.
Figure \ref{t_mat} shows the result at low temperature.
We plot two lines corresponding to $t_\uparrow $ and $t_\downarrow $ for each parameter set. The coincidence of the two indicates that 
the Pad$\acute {\rm e}$ approximation is reliable. 
There appears a peak at the 
Fermi level, which is due to
the underscreened Kondo effect. 
Note that
the spectral shape is far from Lorentzian, 
but is characterized by two different energy scales,
which is most evident in the case of $(J_1,J_2)=(0.2,0.05)$.
This feature is in marked contrast with the spectrum 
in the ordinary Kondo model, which has a single energy scale $T_{\rm K}$.


\section{CEF Singlet vs Kondo Singlet} 
\label{singlet}
We now discuss physical properties in the region (II) where the ground state is a singlet state. 
The competition between the Kondo effect and the CEF effect determines the behavior of this system.

\subsection{Impurity $t$-matrix} \label{sec_t_mat}

In the perturbation theory with respect to $J_1$ and $J_2$, the imaginary part of the $t$-matrix has a threshold singularity 
since the conduction electron whose energy is lower than the effective CEF splitting cannot be scattered by the local state. 

\begin{figure}
\begin{center}
\includegraphics[width=7.5cm]{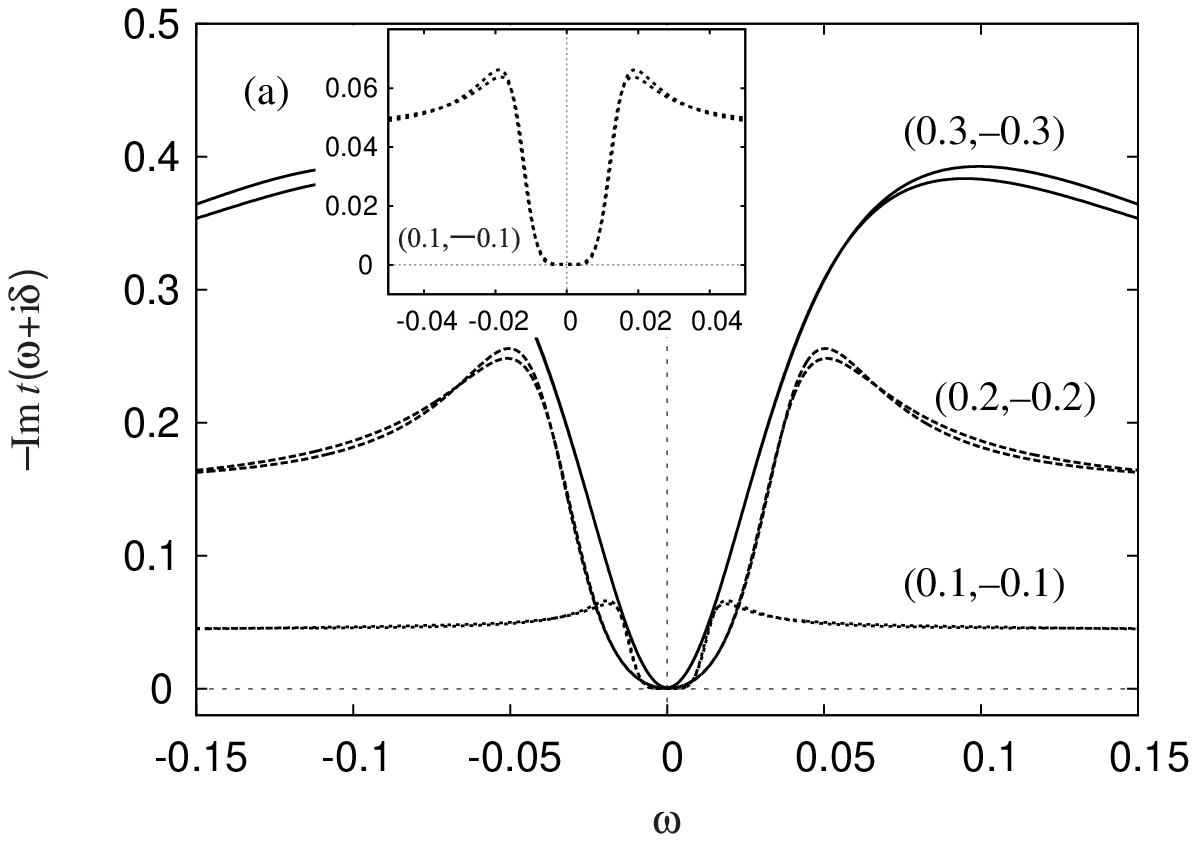}
\includegraphics[width=7.5cm]{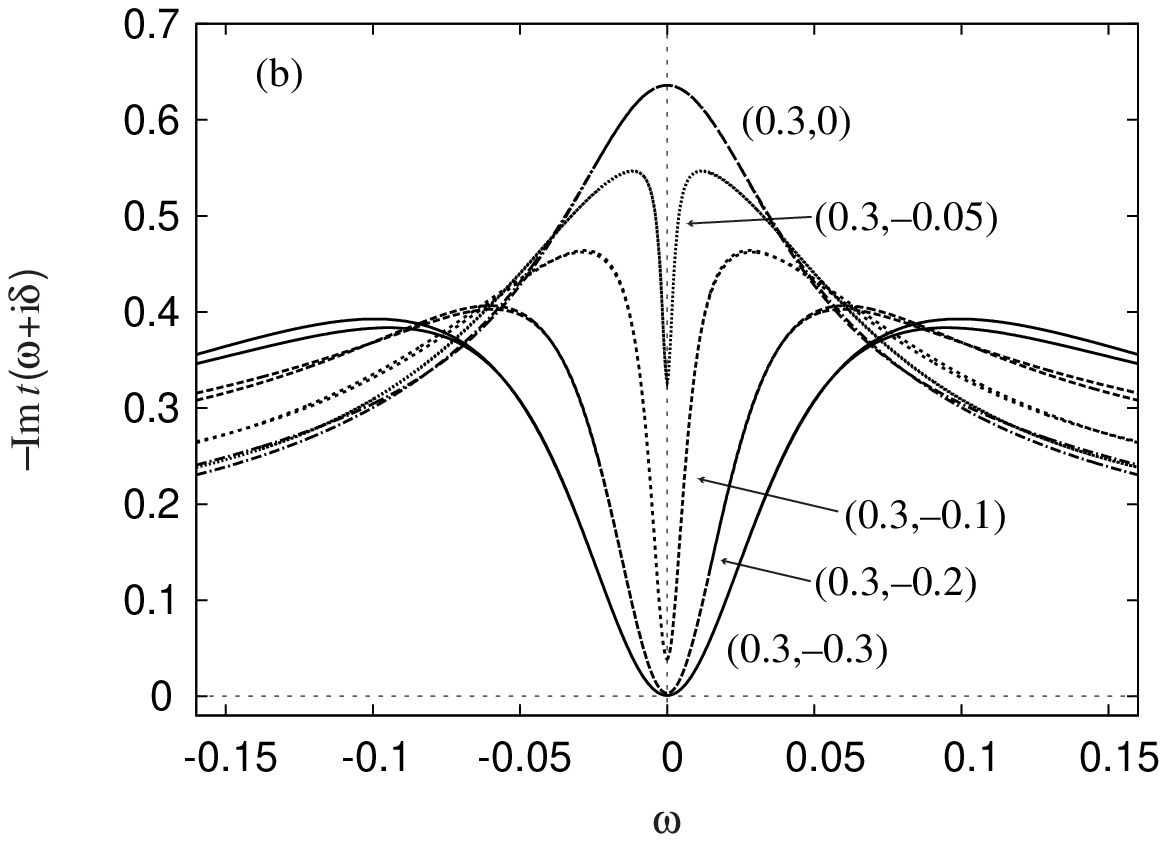}
\end{center}
\caption{The imaginary part of the 
$t$-matrix at $T=0.001$ in the region (II) with the condition (a) $J_1 = -J_2$, and 
(b) $J_1 = 0.3$. 
The two lines for each $(J_1, J_2)$ correspond to up and down spins, and they should coincide in the exact results.
The parameters $(J_1, J_2)$ are shown as (a) triangles, and (b) squares in Fig. \ref{fig2}.}
\label{fig4}
\end{figure}

Figure \ref{fig4} shows numerical results for 
$-\imag t_\sigma (\omega +{\rm i} \delta )$  in the region (II), obtained by the procedure of ref.\citen{bib1}.
Figure \ref{fig4}(a) 
shows the case with a gap structure.
In the case of a small value of $r$ defined by eq.(\ref{ratio}),
the spectral function is similar to the step function. 
This is because the CEF effect is dominant compared to the Kondo effect. 
A typical case with $(J_1, J_2) = (0.1, -0.1)$ ($r \ll 1$) is shown also in the inset of Fig. \ref{fig4}(a).
As $r$ grows, the gap structure becomes obscure.

\begin{figure}
\includegraphics[width=7.5cm]{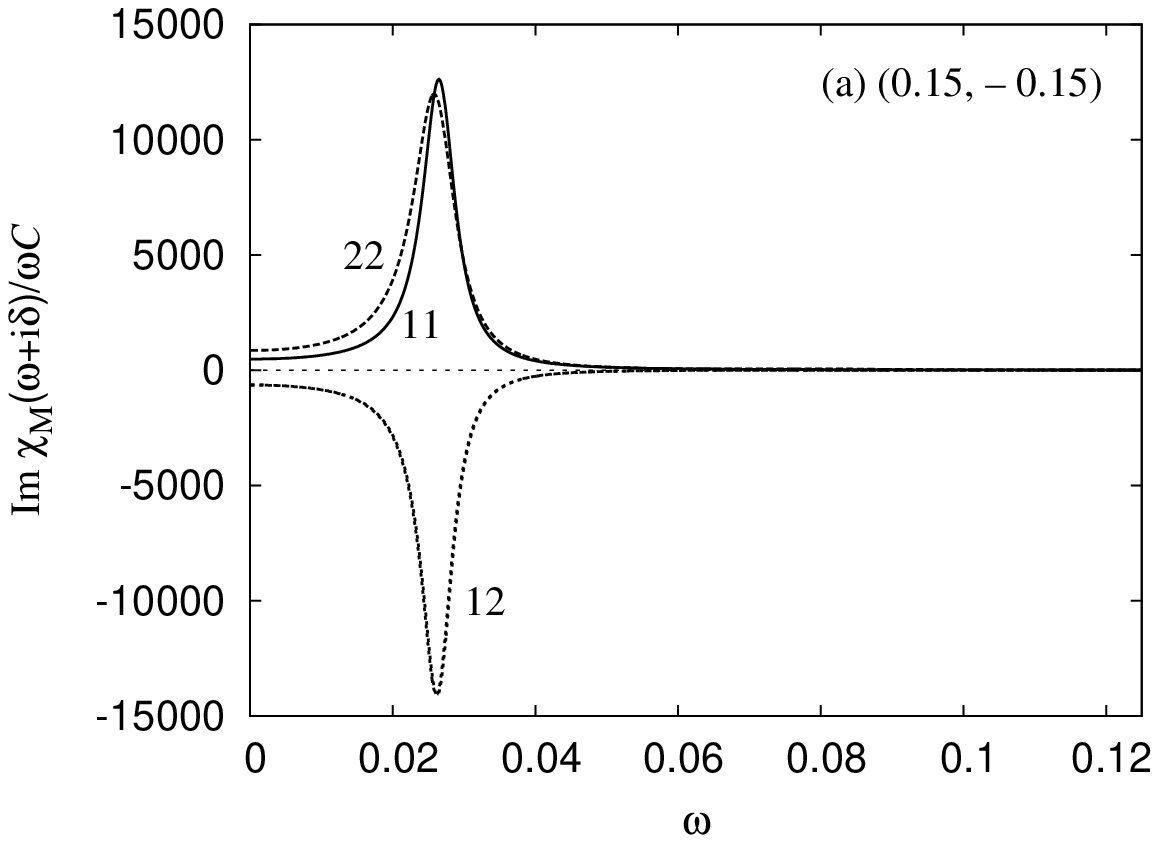} 
\includegraphics[width=7.5cm]{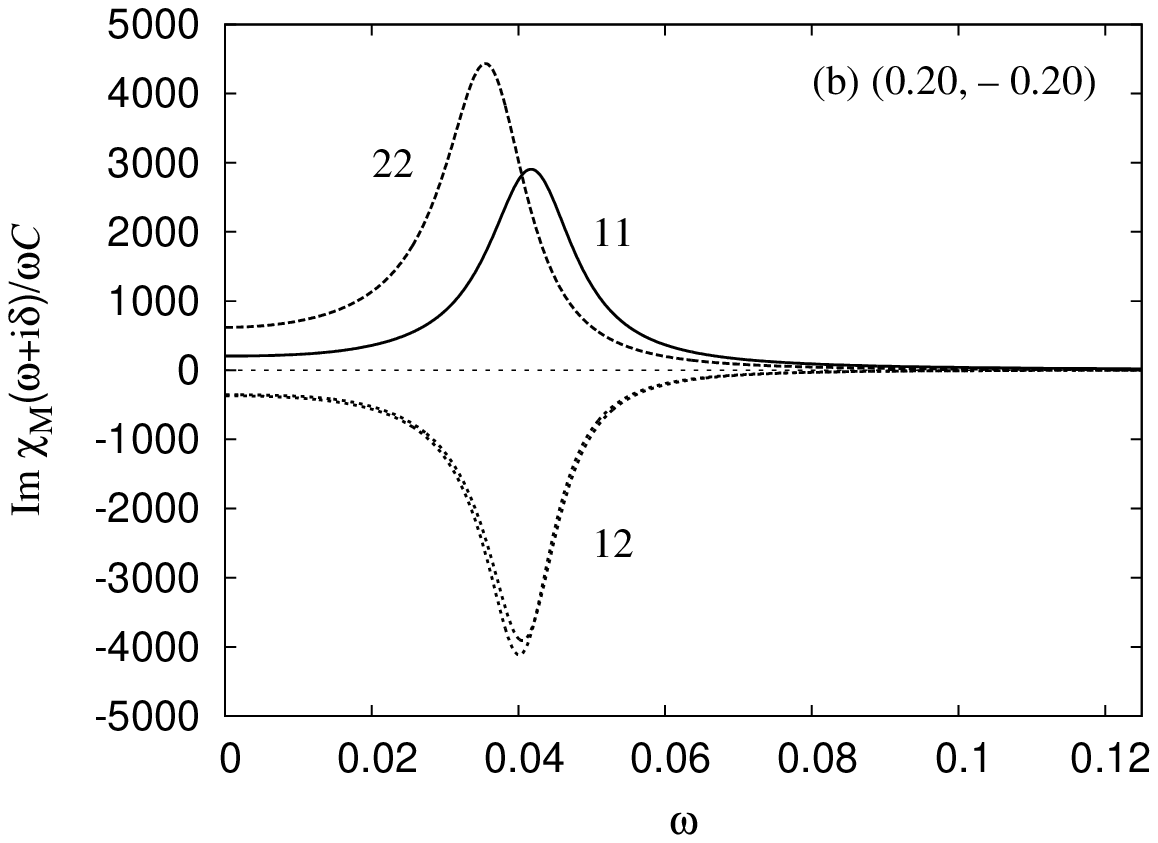} 
\includegraphics[width=7.5cm]{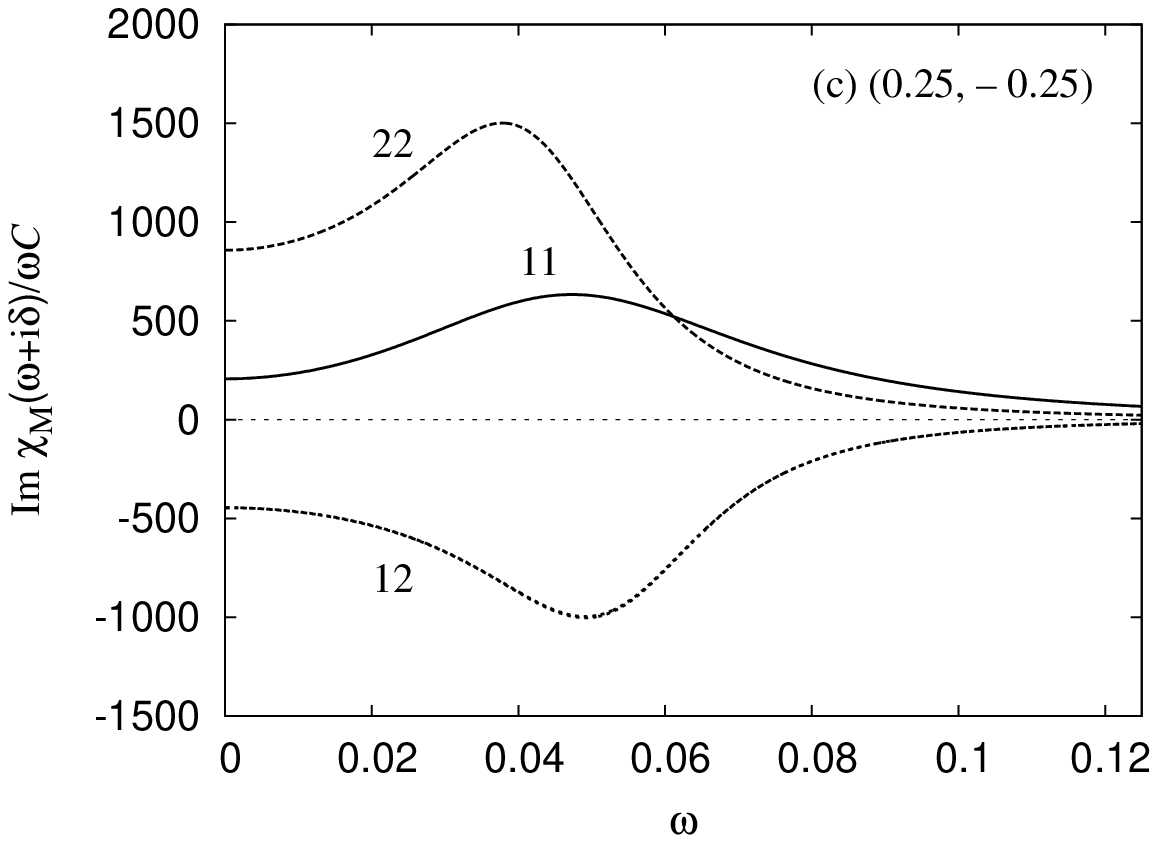} 
\includegraphics[width=7.5cm]{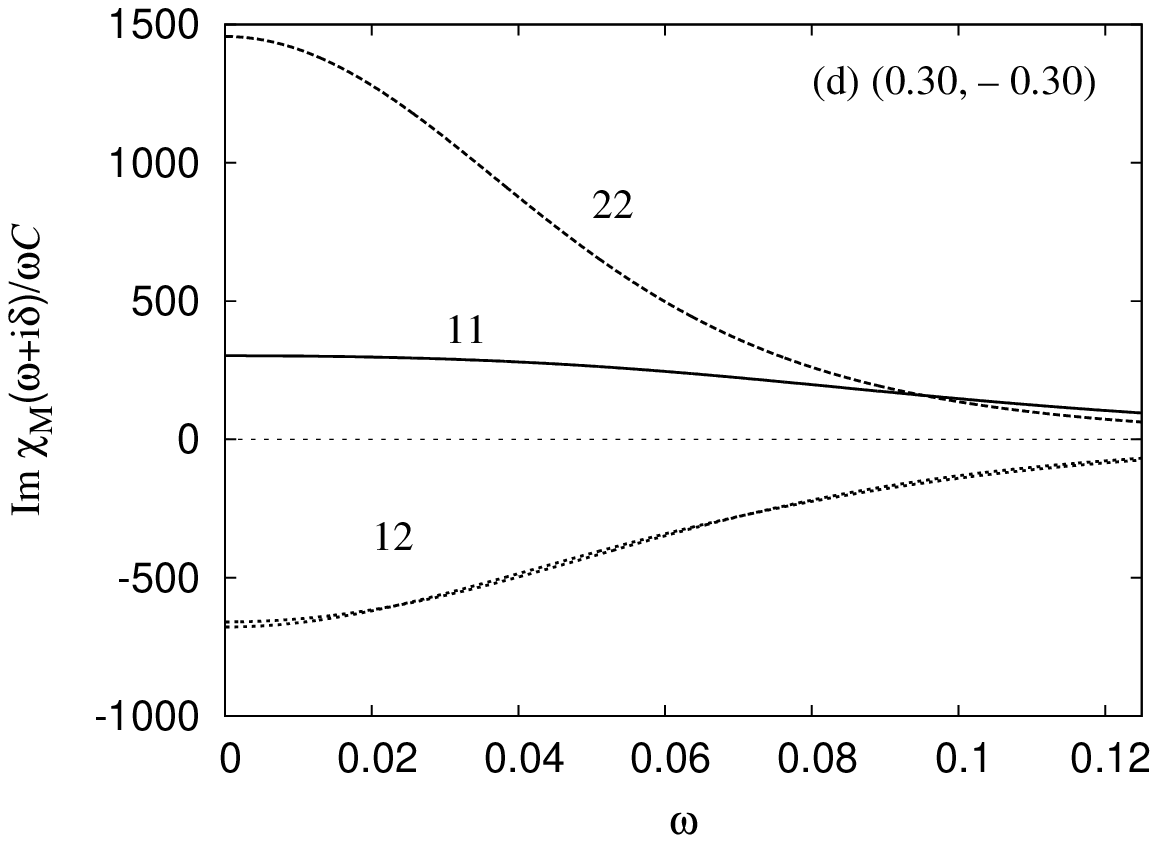}
\caption{
The two-particle spectrum with the condition $J_1=-J_2$ at $T=0.001$. 
The parameters $(J_1, J_2)$  are shown as triangles in Fig. \ref{fig2}.}
\label{fig5}
\end{figure}
We now fix $J_1 = 0.3$ and vary $J_2$ from zero to $-J_1$.
Figure \ref{fig4}(b) shows the results.
With $(J_1, J_2) = (0.3, -0.3)$ ($r \sim 1$), 
for example, the spectral shape has the two broad peaks at $\omega \sim \pm 0.1$.
As $|J_2|$ decreases with $J_1$ fixed, the effective CEF splitting decreases according to eq.(\ref{eqn_delta}) while the Kondo temperature $T_{\rm K}$ defined in eq.(\ref{eqn_kondo}) is constant. 
Hence the ratio $r$ increases with decreasing $|J_2|$. 
The gap in the spectrum becomes narrower as $|J_2|$ decreases, reflecting the smaller effective CEF splitting.  
Note that the temperature $T=0.001$
is not low enough compared to the effective CEF splitting for $J_2 = -0.05$ and $-0.1$. 
Hence the spectrum at the Fermi level is not zero.
In the case of $J_2=0$, which gives $r=\infty$,
these two peaks combine into one. 
The case corresponds to the ordinary Kondo effect, however with an extra free spin.

\subsection{Dynamical Susceptibility} \label{sec_dynamical}
\begin{figure}
\begin{center}
\includegraphics[width=7.5cm]{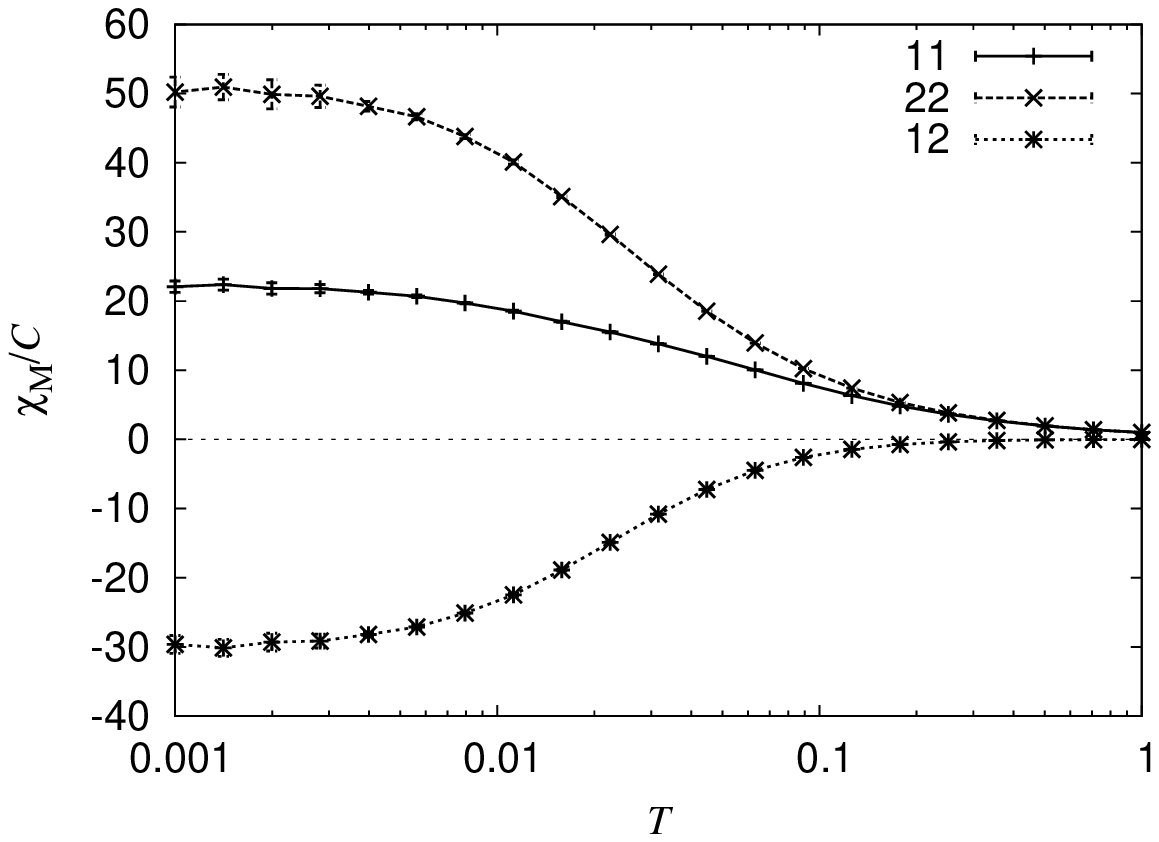}
\end{center}
\caption{Temperature variation of static susceptibilities with $(J_1, J_2) = (0.3, -0.3)$.}
\label{fig6}
\end{figure}
\begin{figure}
\begin{center}
\includegraphics[width=7.5cm]{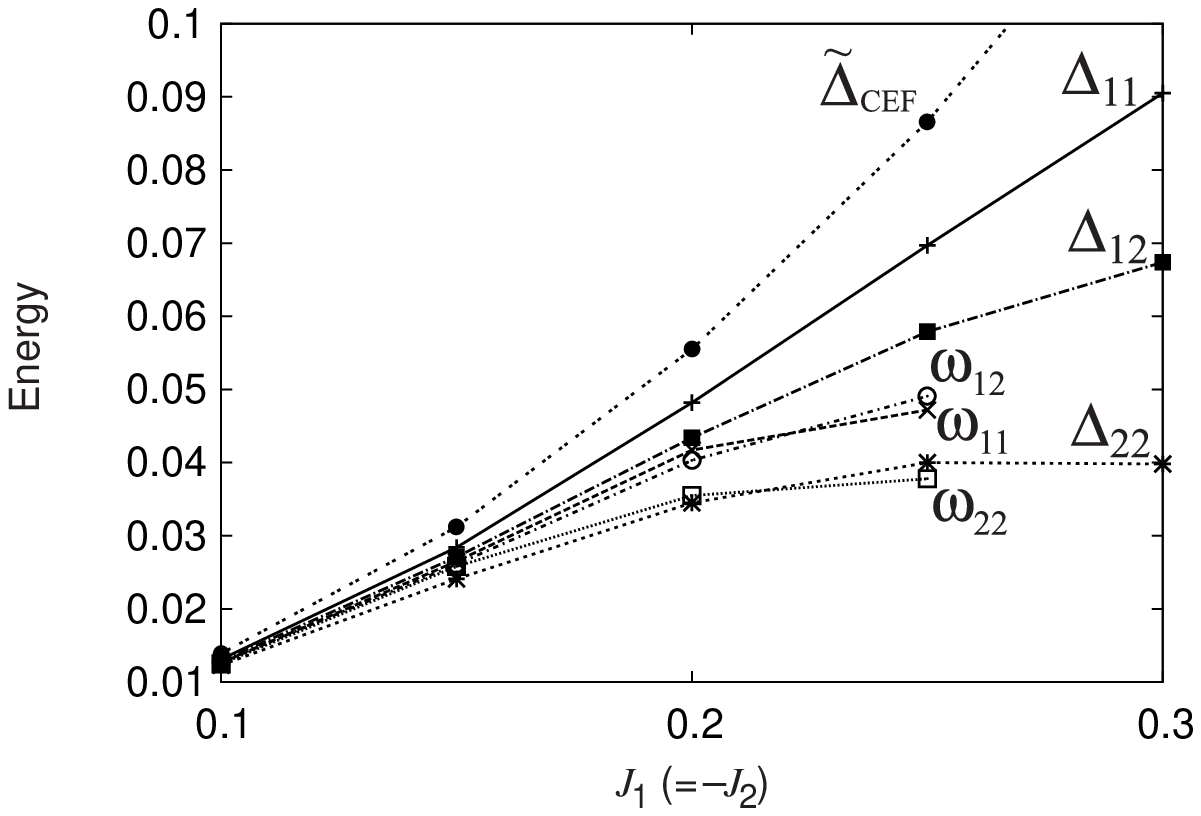}
\end{center}
\caption{
The fitted values of $\Delta _{\mu \nu }, 
\tilde \Delta _{\rm CEF}$ defined by eq.(\ref{eqn_delta}),
and
$\omega _{\mu \nu }$ as a function of 
the coupling constant $J_1=-J_2$. 
In the CEF picture, all these excitation energies should be the same.
The CEF picture becomes reasonable for $J_1 \lesssim 0.15$.
The parameters are shown as triangles in Fig. \ref{fig2}. 
}
\label{fig7}
\end{figure}
We derive the dynamical susceptiblity following the procedure of ref.\citen{bib1}.
Figure \ref{fig5} shows the two-particle spectrum $\imag \chi _{\rm M} ^{\mu \nu}(\omega +{\rm i}\delta )/ \omega C$ at low temperature with $J_1 = -J_2$. 
The susceptibilities should satisfy the relation $\chi _{\rm M} ^{12} = \chi _{\rm M} ^{21}$, which have been calculated separately to check the accuracy of the Monte Carlo simulation.
The corresponding results for 
$\chi _{\rm M} ^{12}$ and $\chi _{\rm M} ^{21}$ 
are close enough to each other.  Hence 
the validitiy of the Pad$\acute {\rm e}$ approximation
is confirmed.
The sign $\imag \chi _{\rm M} ^{12} (\omega) < 0$ corresponds to the negative sign of $\real \chi _{\rm M} ^{12} (0)< 0$ by the Kramers-Kronig relation, and
indicates an antiferromagnetic correlation between the pseudo spins. 
We find an inelastic peak 
for about $J_1 < 0.25$ (see Fig. \ref{fig5}(a)--(c)). 
This peak corresponds to excitation from the CEF singlet to the triplet. On the other hand, 
no inelastic peak is observed at $J_1 = 0.3$ as shown in Fig. \ref{fig5}(d). Instead, the spectrum shows only a broad quasi-elastic peak.

Let us compare the two-particle spectrum with 
the $t$-matrix, which corresponds to the single-particle excitation of the local state. As shown in Fig. \ref{fig4}, the single-particle spectrum has a gap whose shape depends on the strength of the interaction. 
With $r \ll 1$ as shown in Fig.\ref{fig4}(a)$(0.1,-0.1)$ and Fig.\ref{fig5}(a), 
the single-particle spectrum looks like a step function, 
and the two-particle spectrum has the inelastic peak corresponding to the gap. This means that the ordinary CEF picture is valid, 
and the Kondo effect is insignificant. 
On the other hand, with $r \sim 1$ as shown in Fig. \ref{fig4}(a)$(0.3,-0.3)$ and Fig. \ref{fig5}(d), 
the two-particle spectrum has a broad quasi-elastic peak.
We emphasize that the single-particle spectrum still has a clear gap.
Namely, the system with $r \sim 1$ shows both features of the Kondo and CEF effects. 
In this case, the simple CEF picture is no longer valid.

The validity and breakdown of the CEF picture is also seen in the components of the static and dynamical susceptibilities.
In Fig. \ref{fig5}, we have imposed the condition $J_1 = -J_2$, which leads to $I_{\rm t} = 0$. 
In the case of $r \ll 1$ as shown in Fig. \ref{fig5}(a), combination of 
$\chi _{\rm M} ^{\mu\nu}$
according to eq. (\ref{cor_ts}) gives small
$\chi _{\rm t}$.  This is understood because the triplet lies higher than the singlet. 
With stronger coupling as shown in Fig. \ref{fig5}(b)--(d), on the other hand, $\chi _{\rm t}$  is no longler
negligible although the interaction contains only $\mib{X}^{\rm s}$.
This increase $\chi _{\rm t}$ is due to the Kondo effect.

Let us turn to the case of $J_1 < |J_2|$ and $J_1 \rho _{\rm c}\ll 1$, corresponding to the point with asterisk in Fig. \ref{fig2}. 
Here $I_{\rm t}$ is negative and $r\ll 1$.
In this case, $\chi _{\rm t}$ is small, and the inelastic peak can be seen in the two-particle spectrum.
The behavior of this parameter is similar to the case of $(J_1 , J_2) = (0.1, -0.1)$. 
This similarity is understood since a negative $I_{\rm t}$ is renormalized to zero 
at low temperature.
Therefore, we may say that $\chi _{\rm t} \neq 0$ is caused by a large antiferromagnetic coupling $J_1$. 
Owing to this interaction, the ground state of the system changes to more stable Kondo singlet from the CEF singlet. 
As in the ordinary Kondo system,  this singlet ground state in the ST Kondo model is strongly coupled with conduction electrons.

\subsection{Static Susceptibility}
The static susceptibility with $(J_1, J_2) = (0.3, -0.3)
$ is shown in Fig. \ref{fig6}. 
The susceptibility tends to a constant
at low temperature. 
This is related to the van Vleck paramagnetism which arises from virtual transitions to the triplet CEF level. 
We can see this paramagnetism for other parameters in the region (II) such as $(J_1, J_2) = (0.1, -0.1)$.
Note that the paramagnetic behavior can be seen even in the case where there is no inelastic peak in the two-particle spectrum. 

In the singlet-triplet system with the CEF splitting $\Delta $, the van Vleck susceptibility is given by $\chi = 2|\langle {\rm s}|J_z|{\rm t0} \rangle | ^2 / \Delta $, where $| {\rm s} \rangle $ and $| {\rm t}m \rangle$ are the singlet state and the triplet states ($m=+1,0,-1$), respectively. 
To study the van Vleck susceptibility in our model more precisely, 
we define the parameter $\Delta _{\mu \nu }$,
which plays the role of effective CEF splitting,
 as follows:
\begin{eqnarray}
\left. \chi _{\rm M}^{\mu \nu }
\right| _{T=0} = \frac{2}{\Delta _{\mu \nu }}
\langle {\rm s}|S^z_\mu |{\rm t}0 \rangle \langle {\rm t}0|S^z_\nu |{\rm s} \rangle
\label{eqn_van_vleck}
\end{eqnarray}
In the ST Kondo model, the matrix elements in eq. (\ref{eqn_van_vleck}) equals to $(-1)^{\mu + \nu } / 4$. 
Figure \ref{fig7} shows the fitted results for $\Delta _{\mu \nu }$,
together with
the excitation energy $\omega _{\mu \nu }$
corresponding to the inelastic peak in the two-particle spectrum, 
and $\tilde \Delta _{\rm CEF}$ defined in eq.(\ref{eqn_delta}). 
In the weak coupling range with $J_1 \lesssim 0.15$, these values are almost the same. 
Hence, in this case, the effective CEF splitting is close to
$\tilde \Delta _{\rm CEF}$ derived by the second-order perturbation theory.
In other words, the Kondo effect is unimportant. 
This energy has appeared
also in the $t$-matrix shown in Fig. \ref{fig4}(a). 
Namely, the magnitude of the energy gap is nearly equal to $2\tilde \Delta _{\rm CEF}$, 
and the van Vleck paramagnetism accounts for the static susceptibility. 

As the coupling increases, however, such values as $\Delta _{\mu \nu}$ and $ \omega _{\mu \nu}$
deviate from each other. 
Because of the Kondo effect, the CEF picture cannot explain the low-temperature susceptibility in this region.  

\section{Summary} 
\label{summary}
In the present paper, we have extended and applied the CT-QMC method to the ST Kondo model where the conduction electrons interact with two pseudo spins. 
We have derived the impurity $t$-matrix, spin susceptibilities and singlet occupation rate in our model. Since the CT-QMC does not use any approximation, the results given in this paper are highly reliable with only statistical errors.

In the doublet ground-state region ($J_1,J_2 >0$), the behavior of this system is understood as the underscreened Kondo effect. 
In the region $J_1 J_2 < 0$ for the singlet ground-state, the CEF effect competes with the Kondo effect. In the case of $T_{\rm K} \ll {\tilde \Delta} _{\rm CEF}$, the single-particle spectrum has a gap structure at low temperature, and the two-particle spectrum has an inelastic peak corresponding to the gap. 
In the case of $T_{\rm K} \sim {\tilde \Delta} _{\rm CEF}$, however, the inelastic peak vanishes in the two-particle spectrum, or the dynamical susceptibilty.
Instead, a broad quasi-elastic peak appears due to the Kondo effect.
The single-particle spectrum retains a gap even in this regime. 
This contrasting behavior between single- and two-particle spectra results from the competition between the Kondo effect and the CEF effect. 

Using the algorithm given in this paper, it is possible to apply the CT-QMC method to more complicated models such as the quadrupolar Kondo model proposed by Cox.~\cite{bib9} 
As another extension,
we can apply the CT-QMC to the lattice system using dynamical mean-field theory (DMFT), in which the problem is reduced to an effective impurity problem. 
We shall study the ST Kondo lattice system in the future work.

\section*{Acknowledgment}
The authors thanks Dr. A. Kiss and Mr. A. Yamakage for useful discussion in the strong coupling limit, and Dr. H. Yokoyama for his advice in numerical calculation.  
This work was supported by a Grant-in-Aid for Scientific Research No.20340084.

\appendix
\section{CT-QMC for Ferromagnetic interaction}
As noted in ref.\ref{bib1_x}, the CT-QMC for the Coqblin-Schrieffer model is applicable only to the antiferromagnetic coupling ($J>0$) because of the negative sign problem. 
In this section, we show that the algorithm can be applied to the ferromagnetic coupling ($J<0$) by a slight modification.

The exchange interaction term can be written in the following form:
\begin{eqnarray}
J \sum_{\sigma \sigma '} X_{\sigma \sigma '} c_{\sigma '}^\dagger c_\sigma =
J\sum_{\sigma \sigma '} X_{\sigma \sigma '} (c_{\sigma '}^\dagger c_\sigma - \alpha \delta _{\sigma \sigma '}) + \alpha  J \nonumber
\end{eqnarray}
The constant term may be neglected. Here we have introduced the parameter $\alpha $, which gives the equal-time Green function as discussed in ref. \ref{rub}. 
This parameter is set as
\begin{eqnarray}
\alpha = \left\{
\begin{array}{c}
1\  (J>0)\\
0\  (J<0).
\end{array}
\right.  \nonumber
\end{eqnarray}
The two choices give either $g(\tau =+0)$ for $J>0$ or $g(\tau =-0)$ for $J<0$ as the equal-time Green function.
We have found that a simulation using this expression does not encounter the negative sign configuration for $N=2$. 
The absence of the negative sign is understood by considering the $k=1$ term in eq.(\ref{eqn_weigh}).

We note that the procedure
is valid only in the case of $N\leq 2$, where $N$ is the number of the local states.
Namely, the interaction has the operator $X _{\uparrow \downarrow}$ and $X _{\downarrow \uparrow}$ which change the local state. In the $N=2$ case, the total number of these operators in a configuration $q$ must be even. 
Then the contribution to the weight $W(q)$ has 
no difference between the $J>0$ and $J<0$ cases. 
Hence we can deal with the ferromagnetic interaction in the case of $N=2$.  This fortuitous situation does not occur for 
$N\geq 3$.

\label{lastpage}
\newpage

\end{document}